\documentclass[a4paper, 12pt]{article}

\usepackage{graphicx}
\usepackage{dcolumn}
\usepackage{bbm}\usepackage{nicefrac}
\usepackage[latin1]{inputenc}
\usepackage{epsfig}
\usepackage{pifont}
\usepackage{amsmath,amssymb}

\newcommand{\N}{\mathbbm{N}}

\newcommand{\s}{\sigma}
\newcommand{\xipar}{\xi_\parallel}
\renewcommand{\H}{\mathcal{H}}

\begin{document}
 
\title{Radial Distribution Function for Semiflexible Polymers Confined in Microchannels}
 
\author{Patrick Levi and Klaus Mecke\\ Institut f\"ur Theoretische Physik,\\ Universit\"at  Erlangen-N\"urnberg,\\ Staudtstra{\ss}e 7, 91058 Erlangen, Germany}
 
\date{\today}\maketitle

\begin{abstract}An analytic expression is derived for the distribution $G(\vec{R})$ of the 
end-to-end distance $\vec{R}$  
of semiflexible polymers  in external potentials  to elucidate the effect of confinement
 on the mechanical and statistical  properties of biomolecules. 
 For parabolic confinement the result 
is exact whereas for realistic potentials  a self-consistent ansatz is  developed, so that  
$G(\vec{R})$ is given explicitly even  
for hard wall confinement. 
The theoretical result is in excellent quantitative agreement with fluorescence microscopy 
data for actin filaments  confined  in rectangularly shaped microchannels. This allows 
an unambiguous  determination of persistence length $L_P$ and the dependence of statistical 
properties such as Odijk's deflection length $\lambda$ on the channel width $D$.  It is shown that 
neglecting the effect of confinement  leads to a significant overestimation of  
bending rigidities for  filaments. 
\end{abstract}

Bending of  semiflexible macromolecules  such as actin filaments play a crucial role for the 
mechanical properties of cells \cite{bausch06}.  
In general, 
the motion of biopolymers such as DNA or actin filaments  takes place in
 gels or cytoplasm where 
sterical constraints forces a single molecule  to  bend  in addition 
to the bending constantly induced by thermal motion.  
Advances in microfluidic techniques and in the direct visualization of 
actin filaments 
\cite{gittes93,ott93,kaes94,goff02,koester05} make it nowadays 
possible to study experimentally the interplay of  thermal fluctuations and confinement 
in controlled environments.  
Although the dynamics of single biomolecules  can 
be measured 
for almost 20 years 
by fluorescence microscopy in simple confining geometries 
\cite{houseal89,matsumoto92,perkins94,bustamante03} as well as 
in microfabricated porous arrays \cite{han00,nykypanchuk02},   
the theoretical understanding of such constrained  thermal motion is still hindered by the 
fundamental  statistical  problem to treat
thermal activated undulations of macromolecules and steric repulsion from obstacles  
 on the same analytical level. Whereas thermal bending modes can be treated well in Fourier space 
and steric constraints in real space, respectively,  
the combined interaction of thermal bending in confinement is not solved yet in a satisfactory way.

In order to characterize  thermal fluctuations of polymers of length $L$ one may calculate 
the tangent-tangent correlation function or the radial distribution function, i.e.,  
the probability distribution $G(\vec{r})$ for the distance vector $\vec{r}$ of the two filament  
ends \cite{koester05}.  Because 
end-to-end distances can be measured  easily by labelling the ends of the macromolecule 
the latter served in the past as an important experimental tool to elucidate the physical 
properties of biopolymers including DNA in nanochannels \cite{tegenfeldt04}. 
However, estimated values for bending rigidity $\kappa$ and persistence 
length $L_P=\kappa/(k_BT)$, 
for instance, 
depend on the details of confinement. 
An expression for $G(\vec{r})$ for freely fluctuating single, semiflexible  polymers can be found 
in Ref. \cite{wilhelm96}, 
which is unfortunately not applicable for confined and interacting filaments. 
The effect of confinement on the radial distribution function $G(\vec{r})$ is also  relevant  for the 
microrheology of actin solutions where entanglement plays an important role \cite{gardel03}.  
Thus, an analytical  result for the dependence of $G(\vec{r})$ on confinement 
would provide an important  tool to analyze 
experimental data in microfluidic devices in order to determine  unambiguously the  
physical properties of biomolecules.  
The statistical mechanics of unconstraint fluctuations of 
semiflexible chain molecules ($L<L_P$) is well understood. 
A remarkable successful description  is the worm-like chain 
  model based on  a fluctuating, elastic string \cite{kratky49}.  
However,  
not even the case of parabolic confinement was solved yet beyond tangent 
correlations functions \cite{odijk83}. Here, we present an analytic solution 
for the end-to-end 
distribution function  for strong confinement with $D<<L$ which is in quantitative agreement 
with  experimental findings measured 
in Refs. \cite{goff02,koester05,tegenfeldt04}, for instance.  
Moreover, a self-consistent ansatz allows the unambiguous mapping of non-parabolic 
confinements like hard walls or even van-der-Waals potentials on a quadratic 
Hamiltonian \cite{mcg03}, so that our analytic result is even  applicable on realistic 
substrate-filament interactions.  
This self-consistent ansatz was already successful in predicting the fluctuation spectrum 
of bounded membranes in excellent agreement with experimental 
data \cite{mcg03} as well as for thickness dependence of wetting layers \cite{vorberg01}. 
Here, we present the first application on one-dimensional 
fluctuating filaments  
 and compare our analytical result with experimental data for actin filaments \cite{koester05}.

Actin filaments have diameters of $\approx 8$nm, lengths of  $L=11-13\mu$m  
and a persistence length of about $L_P \approx 15\mu$m yielding thermal fluctuations which 
can be observed by 
optical fluorescence microscopy \cite{koester05}.  
Thus,  in addition to the importance of these thermal fluctuations for the 
biological functioning of cells, actin filaments  can be used as an experimentally 
accessible  model system for semiflexible polymers where $L<L_P$.    
The statistical properties of  confined filaments depend crucially on the ratio $L/\lambda$ 
of the total length $L$ of the filament to Odijks deflection length $\lambda$ which equals  
approximately the number of bendings caused by the walls. 
Because the channel width $D$ in the 
experiments described in Ref. \cite{koester05} is much smaller than the length 
of the macromolecules  we focus here on the strong 
confinement limit $D<<L$. 
\begin{figure} 

\includegraphics[scale=0.54, angle=0]{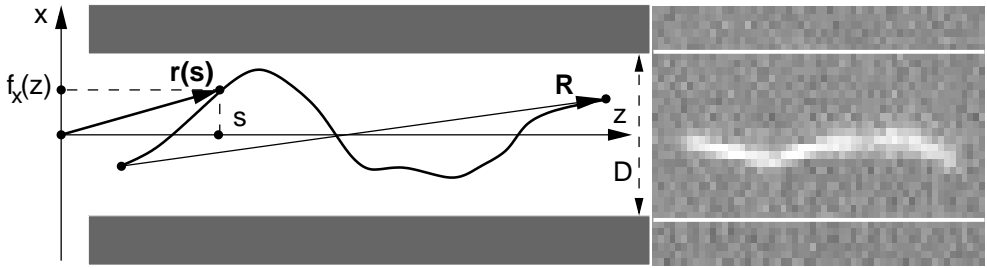}
\caption{Microfluidic devices fabricated by soft photo-lithography makes it possible 
to observe thermal fluctuations of actin filaments in confining microchannels 
by fluorescence microscopy (image scanned from Ref. \cite{koester05}).  
A theoretical description can be 
based on a Monge-parametrization of the perpendicular deviation $f(z)$ of the filament 
from the $z$-axis 
along the channel instead of its arc length $s$.  
Despite the limited spatial resolution of the image the end-to-end distance 
$\vec{R}$ can be measured accurately which makes its distribution $G(\vec{R})$ an 
ideal quantity to determine physical parameters of the filament.   $G(\vec{R})$ is 
given analytically in Eq. (\protect\ref{ete})  in very good agreement with experimental data 
shown in Fig. \ref{fig:experiment}. 
 } \label{fig:kanal}
\end{figure}
An actin filament of length $L$ is modelled as  a differentiable curve in space parametrized 
by the position vector $\vec{r}(s)$ and the arc length $s\in [0,L]$ 
(see Fig. \ref{fig:kanal}). 
Because  polymers 
in  microchannels are elongated as shown in the fluorescence microscopy image, 
one may ignore loops or overhangs of $\vec{r}(s)$. Applying a Monge parametrisation 
one can write 
 the  thermally 
induced deviation $\vec{f}(z)=(f_x(z),f_y(z))$ from the z-axis $\vec{e}_z$ parallel to  
the channel walls as  single-valued functions, i.e., 
$\vec{r}(s)=\left(\vec{f}(z),\; z\right)$.  
Then, the Hamiltonian of an elastic string with bending rigidity $\kappa$ can be written as 
\begin{equation}
\label{hamiltonian}
\H_\text{bend} = \frac{\kappa}{2}\int_0^{L}ds 
\left( \frac{\partial \vec{t}}{\partial s}\right)^2 
= \frac{\kappa}{2}\int_0^{L_z} {dz \; \left( \partial_z \vec{t}(z)\right)^2 \over \sqrt{1+(\partial_z \vec{f})^2}}
\end{equation} 
with the 
normalized tangent vector $\vec{t}(s)={\partial \vec{r}\over \partial s}    
= \left(\partial_z \vec{f}, 1\right)/\sqrt{1+(\partial_z \vec{f})^2}$, the abbreviation 
$\partial_z={\partial \over \partial z}$,  and the 
projection $L_z$   of the contour length $L$ on the $z$-axis of the channel.  
Notice, that $L_z(\vec{f})$ itself is a function of $\vec{f}$ given by the implicit relation 
$L=\int_0^{L_z(\vec{f})}dz\;\sqrt{1+(\partial_z\vec{f}(z))^2}$ at fixed contour length $L$.  
The channel walls are assumed to be purely repulsive with a quadratic cross-section and width $D$. 
To model the  confinement we  add an energy term 
$\H_\text{pot}=\frac{1}{2}\int_0^{L_z}dz\; U(\vec{f}(z))$ 
where the potential $U$ describes the interaction of the filament with the substrate material.

In order to avoid non-Gaussian path integrals in the beginning, we start  with a parabolic potential 
$U(\vec{f})=\frac{E}{2}\vec{f}^2$ perpendicular to the channel axis $\vec{e}_z$. In a second step 
we map self-consistently non-Gaussian potentials on an effective parabolic potential strength $E$ 
and derive an explicit expression for the dependence of $E$  on the channel width $D$. 
Accordingly, we expand also ${\cal H}_{bend}$ up to second order in $\vec{f}$ because the 
fluctuations $f_{x,y}(z)$ are expected to be small at room  and even physiological temperatures, 
yielding 
$\H = \H_\text{bend}+\H_\text{pot}=\frac{1}{2}\int_0^{L_z}dz\; \left[ \kappa (\partial_z \vec{f})^2 +E \vec{f}^2\right]$. Thus, the statistical properties of the actin filament are determined by only two characteristic length scales, the persistence length $L_P=\kappa/(k_BT)$ and the 
deflection length $\lambda = \xipar =\sqrt[4]{\frac{4\kappa}{E}}$ introduced by Odijk \cite{odijk83}. 
Whereas $L_P$ is a quantitative measure for the thermal flexibility of the filament, 
$\lambda$ is the characteristic length of bendings forced by the confinement. 
The Gaussian approximation is additionally justified by the channel confinement which suppresses 
effectively 
large fluctuations on length scale larger than $\lambda$. 
Notice, that the  inextensibility constraint $|\vec{t}|=1$  is violated in the  Gaussian 
approximation, although  automatically fulfilled 
by using the arc length $s$ in $\vec{t}={\partial \vec{r}\over \partial s}$.

We exploit open end boundary conditions and mirror the polymer at the origin 
in order to be able to 
apply periodic boundary conditions for this augmented configuration. 
The mirroring of the polymer suppresses sine-modes but doubles the cosine-modes 
by allowing  wavevectors ${\pi \over L}k$ ($k\in\N$) 
instead of ${2\pi \over L}k$ as for usual periodic boundary conditions. 
By doing so, the end and the midpoint of the periodic string, i.e. the two endpoints 
of the original filament can fluctuate freely and are not constrained on the same 
excursion, which is crucial for the following analytic calculation. 
This boundary condition is equivalent to the one used by Wilhelm and 
Frey for unconfined polymers \cite{wilhelm96}, so that we recover their result for $E=0$ 
as one can see below.  
Then, $\vec{f}(z)$ can be expressed  
in a discrete Fourier 
sum $f_{x,y}(z)=\sum_{k=1}^\infty \tilde{f}(k)\cos\left({\pi\over L_z} kz \right)$ with 
integer $k$. 
The end-to-end vector $\vec{R}=\int_0^{L}ds\; \vec{t}(s)$

reads in Gaussian approximation in terms of the Fourier amplitudes 
\begin{equation}
\vec{R}=L \vec{e}_z 
-  \sum\limits_{k=1}^\infty \left(\begin{matrix} (1-(-1)^k)\tilde{f}(k) \cr
 L k^2   \tilde{f}(k)^2 /4  \end{matrix} \right) 
   \;+\; {\cal O}(\tilde{f}(k)^4) 
\end{equation}
yielding  the end-to-end distribution function 
$G(\vec{r})=\left\langle\delta \left(\vec{r}-\vec{R}\right)\right\rangle 
=\delta (r_x)\delta (r_y) G_0(r)$ where the angle brackets indicate an ensemble average 
determined by the Hamiltonian \eqref{hamiltonian} in Gaussian approximation. 
Applying Fourier expansion of $\delta \left(\vec{r}-\vec{R}\right)$ as well as residuum theorem 
one finds 
\begin{eqnarray}\label{ete}
G_0(r) &=& \frac{1}{L\mathcal{N}}\sum\limits _{k=1}^\infty e^{F(k)(r/L-1)}F(k)\prod\limits _{k'\in\mathbbm{N}\setminus\{k\} }\frac{F(k')}{F(k')-F(k)}  \cr 
 & =& \frac{L_P}{\mathcal{N}L^2}\sum\limits_{k=1}^\infty e^{F(k)({r\over L}-1)} 
 \frac{\pi^2k (-1)^{k-1}}{\sin\frac{\pi k_c^2}{k}}\left(1-\frac{k_c^4}{k^4}\right) 
\end{eqnarray}
with  the dimensionless length ratios  
\begin{equation}
k_c^4= {4\over \pi^4} {L^4\over \lambda^4}  \quad  {\rm and} \quad 
F(k)= {L_P\over L }\pi^2 k^2\left(1+ \frac{k_c^4}{k^4}\right)  \; .
\end{equation} 
An essential step in the derivation of this analytic result is the splitting of 
the product 
\begin{eqnarray}
\label{product} 
\prod\limits _{k'\ne k}\frac{F(k')}{F(k')-F(k)} & = & 
\prod\limits _{k'\ne k} \frac{k'^2}{k'^2-k^2}\prod\limits _{k'\ne k} \frac{1+\frac{k_c^4}{k'^4}}{1-\frac{k_c^4}{k'^2k^2}}
\end{eqnarray}
into  two factors.  The first product equals $(-1)^{k-1}$ whereas the second one can be 
rewritten in terms 
of the functions $p_n(x)=\prod\limits _{j=1}^\infty \left(1+\frac{x}{j^4}\right)$ with  
$p_2(x)=\frac{\sin \left(\pi\sqrt{-x}\right)}{\pi\sqrt{-x}}$. 
Because $p_4(k_c^4)$ is  a numerical factor independent 
on $k$, it can be put into the normalization constant $\mathcal{N}$ which is determined by 
$\int d^3r\; G(r)=1$.

\begin{figure} 
\includegraphics[scale=0.35, angle=270]{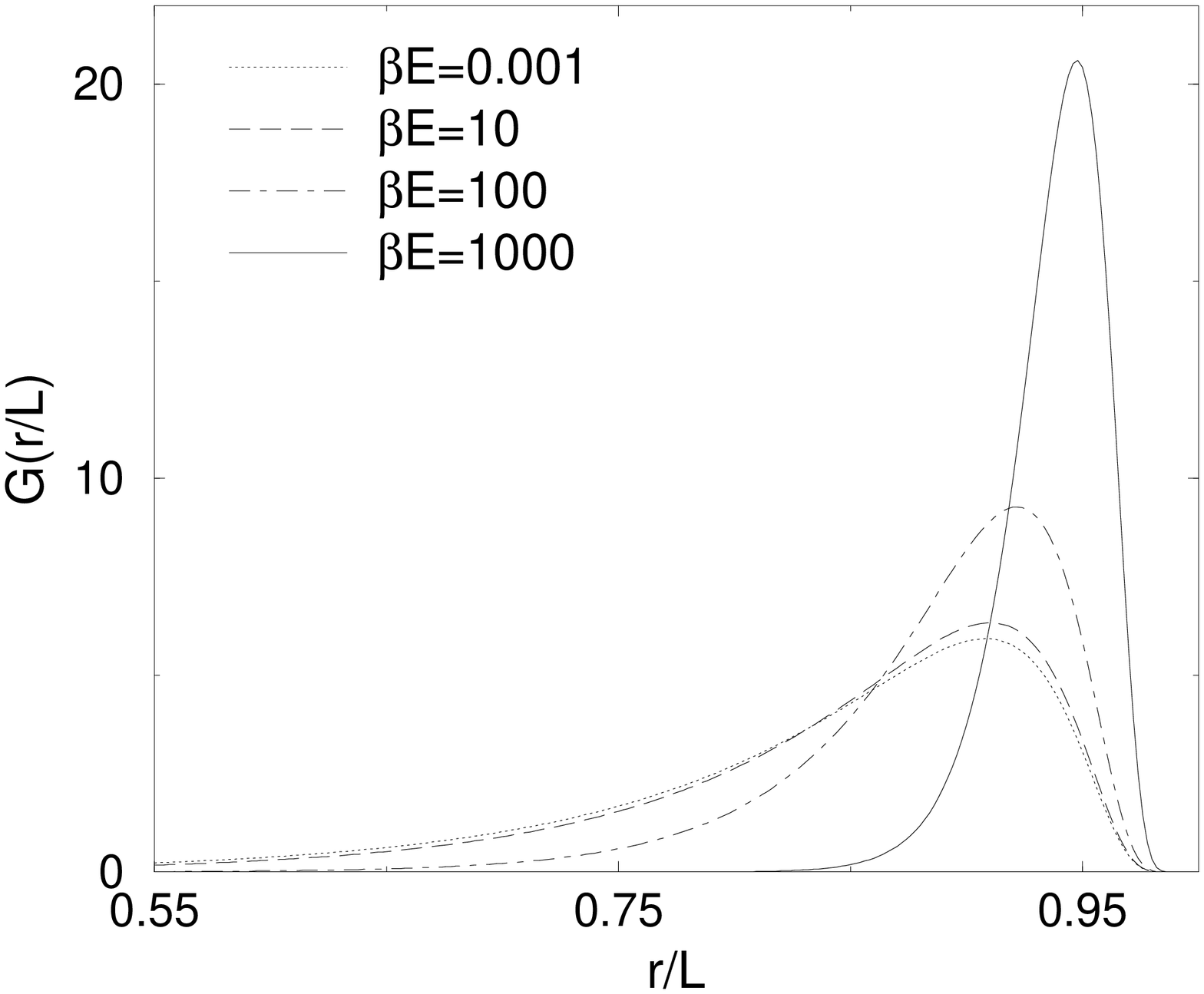} 
\includegraphics[scale=0.35, angle=270]{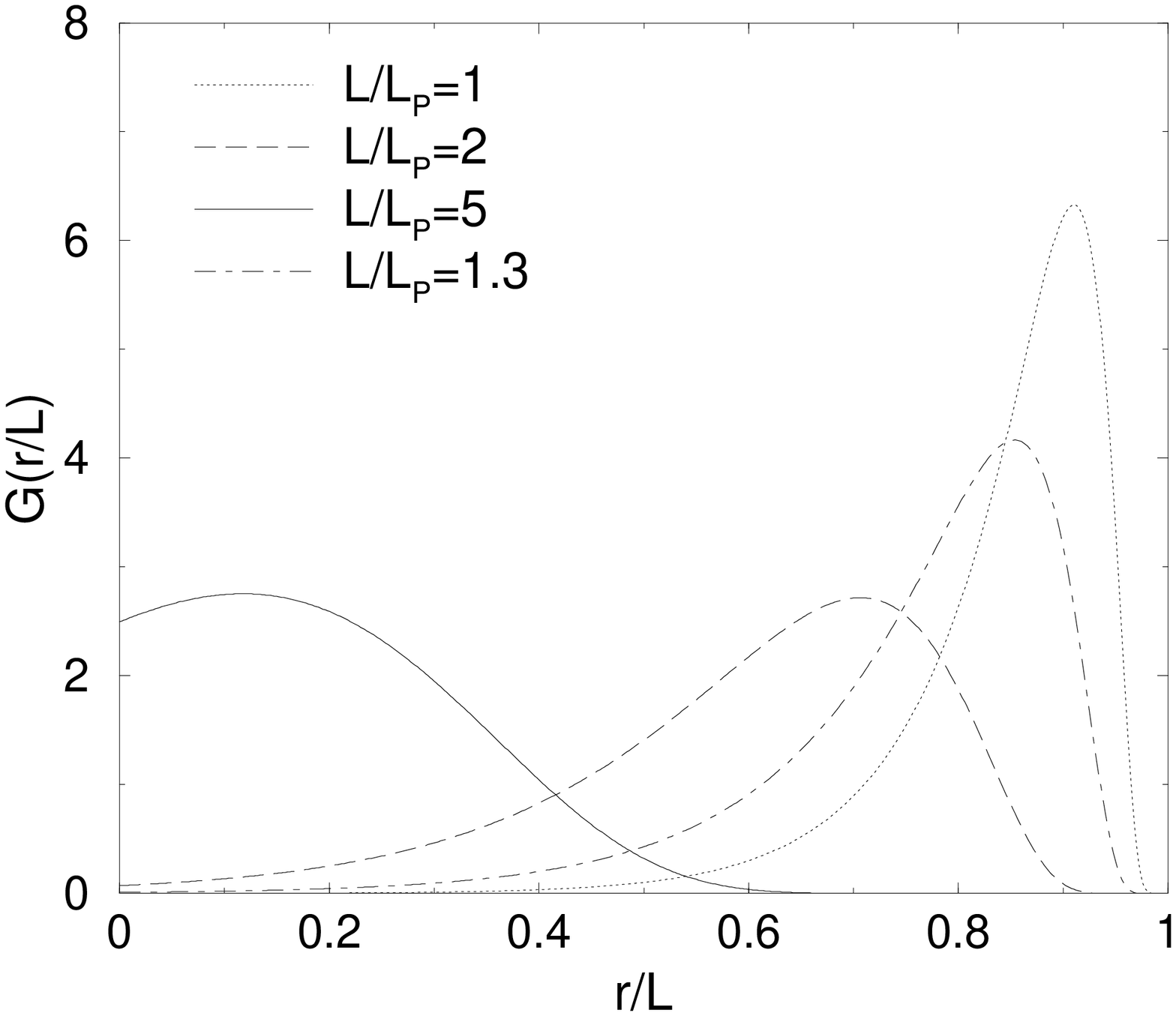}
\caption{Radial distribution function $G_0(R)$ for filaments of   
persistence length $L_P$ as function of the confining potential strength 
$E$ and the  contour length $L$.   
} \label{fig:various_potential}
\end{figure}

In Fig. \ref{fig:various_potential} we illustrate the dependence of $G_0(r)$  on 
the potential strength $E$ as well as on the bending rigidity $\kappa$, i.e., 
on $\lambda$ and $L_P$, respectively. We observe that with increasing $E$ the most 
probable end-to-end distance  $r/L$, i.e., the position of the peak in $G_0(r)$ 
is shifted to higher values   and that the peak gets sharper. This is obvious since strong 
confinement supports  rather straight conformations and short end-to-end distances 
get more unlikely.  
In the limit of vanishing confinement $E\to 0$ one finds a diverging deflection length 
$\lambda \rightarrow \infty$ 
and a vanishing critical wavevector $k_c\rightarrow 0$. In this limit, Eq. \eqref{ete} is 
identical to the  result of Wilhelm and Frey derived for freely fluctuating  semiflexible 
polymers \cite{wilhelm96} where large wavevectors $k$ do not contribute significantly 
to $G_0(r)$ due to bending energies. 
However, the channel walls induce an apparent larger stiffness 
as shown in  Fig. \ref{fig:various_potential}. Thus,  
analyzing experimental data with the result presented in Ref. \cite{wilhelm96} would 
lead to 
significant larger values for the persistence length of the polymer. 
  Finite values of the potential strength $E$, 
i.e., of $k_c$   effectively suppress additionally wavevectors with $k<k_c$ in the radial 
distribution function, so that only modes with $k \approx k_c$ contribute significantly 
to the partition sum.

In order to compare the explicit expression for the radial distribution $G(r)$ given 
by Eq. \eqref{ete} to experimental data we need to relate the average projection length $<L_z>$ 
to the contour length $L$. In Gaussian approximation on finds $<L_z> = L/(1+\sigma'^2)$ with 
the variance 
\begin{equation}
\sigma '^2= <(\partial_z f_{x,y})^2>=\int\limits _0^{q_\text{max}}\frac{dk}{\pi}\frac{k_BT k^2}{E+\kappa k^4}  
\end{equation}
of the first derivative of the filament position $\vec{f}$ with respect to the lateral position $z$. 
The upper border of the integral $\sigma '^2$ $q_\text{max}=2\pi/\xi$ where $\xi$ denotes the length scale where thermal fluctuations disappear, which is either the monomer length of the considered protein or the spatial resolution of the experimental setup as in our case. 
 Evaluating data from microscopy experiments one does  not measure the contour length $L$ of 
the filament but only the projection $L_{||}$ of $L$ on the focal plane. Analogous to $L_z$ 
one finds in Gaussian approximation for the averages the relation  
$<L_z>=<L_{||}>/(1+\sigma '^2/2)$ which is used in  Eq. (\ref{ete}) when comparing 
experimental data with the theoretical result. Notice, that this correction remedies the 
violation of the inextensibility constraint 
$|\vec{t}(s)|=1$ within  Gaussian approximation.

\begin{figure} 
\includegraphics[scale=0.5, angle=270]{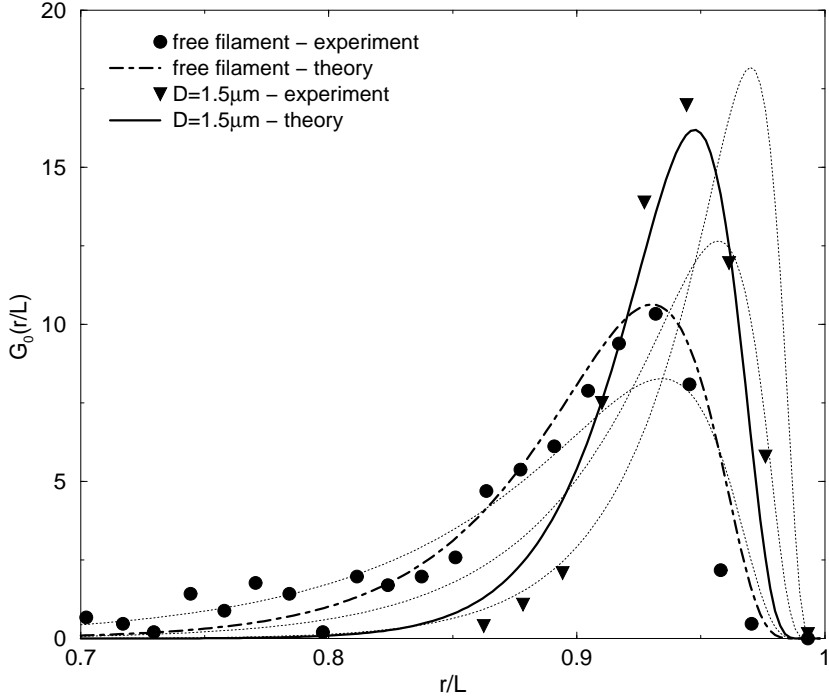}
\caption{The  analytical result for the radial distribution function $G_0(r)$ 
given by Eq. (\protect\ref{ete}) 
(solid line)  describes accurately the  measurements of  the end-to-end distance 
 of the filaments by 
fluorescence microscopy (symbols, see Ref. \protect\cite{koester05}). 
We find consistently the persistence length $L_p\approx 15\pm 3\mu$ for all 
measurements whereas a theory neglecting 
confinement (dotted lines with $E=0$ and $\kappa = 19$, $\kappa = 25$, $\kappa = 35$) 
does not agree with the experimental data. Moreover, comparing the data and the 
theory without confinement ($E=0$) would imply a dependence of the persistence length 
on the channel width $D$.     
} \label{fig:experiment}
\end{figure}

In Fig.  \ref{fig:experiment} we compare the theoretical expression \eqref{ete} to 
experimental distributions measured by  fluorescence microscopy  
from a single actin filament in a microchannel.  
The data and description of the experimental setup can be found in 
Ref. \cite{koester05}. The widths of the channels range from  $D=1.5\mu$m over  
$D=5.8\mu$m to almost infinite widths  where the filament lengths are in 
general larger than $10\mu$m.   We fit the end-to-end distribution function \eqref{ete} 
using 
the potential strength $E$ and the bending rigidity $\kappa$ as free parameters and find 
quantitative agreement between the theoretical prediction for confinement and 
the experimental data.  
In all cases we find consistently  
the persistence length $L_P\approx 15\pm 3\mu$m in good agreement 
with $L_P=18\mu$m from Ref. \cite{koester05} 
considering the difficulties in measuring persistence lengths. 
It is important to notice that the agreement between theory and experiment for different 
channel widths is found for identical persistence length $L_P$. Without taking the effect of 
confinement into account, i.e., a finite value for $k_c$ in Eq. (\ref{ete}) one would 
obtain significant different values for $L_P$.   
Fig. \ref{fig:experiment}(a) shows also that an unconfined filament ($E=0$) cannot be used to 
describe the data. Taking the height of the peak would systematically overestimate the 
persistence length and lead to unrealistic values. 
The quality of the fits with finite $E$ indicates that 
an effective parabolic potential is appropriate, if the strength $E$ 
depends on the channel geometry. Here the channel geometry is set by a 
rectangular cross-section with  
width $D_1=D$ and height $D_2 \approx 1.4\mu$m.  
In a final step we derive an analytic expression to map the channel geometry on a 
parabolic potential with $E(D_1, D_2)$ which can be used in Eq. \eqref{ete} even for 
non-parabolic confinement of the filament.

For hard wall confinement in a channel with rectangular cross-section the potential 
reads $U^{(hw)}(\vec{f}) = U_x(f_x)U_y(f_y)$ with $U_i(f_i)=0$ for 
$-\frac{D_{i}}{2}<f <\frac{D_{i}}{2}$ and $U_i(f_i)=\infty$ otherwise. 
 To do the mapping we employ a technique from Ref. \cite{mcg03} which proofed to work well 
for two dimensional membranes and thin films \cite{vorberg01}. 
It considers the root mean square  fluctuations amplitude 
$\sigma_{x,y}^2=\langle f_{x,y}^2\rangle$ given by 
$\sigma ^{-2}= 2^{3/2}\beta \kappa^{1/4}E^{3/4}$ for parabolic confinement. 

The central idea is to calculate $\sigma ^2$ separately in Fourier space for 
short wavelength fluctuations and in real space for long wavelengths, respectively, and assuming  that both approaches lead to the same value. In real space the polymer is to this end devided into independently fluctuating segments of size $\xipar$. This self-consistent ansatz leads to the implicit equation
\begin{equation}\label{selfcon}
0 = \int_{-\infty}^\infty df\; (f^2-\s_{x,y}^2) e^{ -\beta U_{x,y}(f)\xi_\parallel 
- \frac{f^2}{2\s_{x,y} ^2}}  \; .
\end{equation} 
for the fluctuation amplitude $\s_{x,y}$.
Here, $\xipar = \lambda = \sqrt[4]{\frac{4\kappa}{E}}$ is the correlation length parallel 
to the channel  given by the deflection length $\lambda$. 
In case of the hard wall potential $U^{(hw)}$ we arrive at the  equation 
$\exp\left( -1/(16\mu) \right)= \sqrt{\pi \mu} {\rm erf}\left(1/\sqrt{16\mu}\right)$ for the ratio  
$\mu=\s^2/D^2$. Because $\mu \approx  0.063777$ is constant, one finds 
the scaling 
$\sigma \varpropto D$ analogous to Odijk's scaling relation 
$\lambda\varpropto L_P^{1/3}D^{2/3}$ derived for spherical channels \cite{odijk83}. 
Moreover, one can determine the prefactor analytically by applying   
$4\beta E L_P^{1\over 3} \mu^{4\over 3}D^{8\over 3}=1$  yielding   
the expression 
\begin{equation}
\lambda =\sqrt[4]{\frac{4\kappa }{E}}= 2L_P^{1\over 3}\sigma^{2\over 3} 
= 2 \mu^{1\over 3} \; L_P^{1/3}D^{2/3} 
\end{equation}
for the deflection length. 
Notice, that for a one dimensional filament the ratio $\mu = \sigma^2/D^2 \approx 0.063777$ 
is much smaller compared to the value $\mu\approx 0.2$ for a two dimensional 
membrane \cite{mcg03}.\\
Computing  $E$ for the parameters  of 
the measured data set shown in Fig. \ref{fig:experiment} we find a
good agreement with the experimental data,  which makes our explicit result an 
important tool to analyse biophysical in vitro studies where  more complex constraints  
are relevant due to the biological  environments. 
For instance, spatial constraints can  not only be induced by solid channel walls but also 
by forces between  parallel filaments within the channel. 
Repulsive forces lead to additional extension  whereas 
attractive interactions may cause the formation of bundles of 
semiflexible filaments. 
Our analytical self-consistent approach seems to exhibit  
a discontinuous unbinding transitions of bundles recently found in 
Monte-Carlo simulations \cite{kierfeld05}.   
Also in steady shear flow  the conformational dynamics of individual polymer molecules 
can be visualised and the probability distribution for the polymer extension can be 
measured \cite{smith99,schroeder03}. The explicit expression (\ref{ete}) for the 
radial distribution function  may be applied  to distinguish between extensions due to 
shear flow or 
due to spatial confinement in channels.  
Recently, the fluctuations of semiflexible polymers were visualised in a nematic 
liquid crystal where  confinement due to rod-like colloidal particles leads to 
an elongation of the filament \cite{dogic04}. The observed coil-rod transition 
makes our approach applicable so that the explicit expression for the radial 
distribution function may be used to describe the elongation of the filaments 
beyond the tangent correlation function used in Ref. \cite{dogic04}. 
We expect that our main result, the explicit expression for $G(\vec{R})$, can be used 
to analyze experiments of more complex situations such as solutions of filaments 
even  in inhomogeneous channel geometries  which is crucial for the understanding of 
the cytoskeleton in vivo.

\end{document}